\documentclass[aps,preprint]{revtex4}%
\usepackage{amsfonts}
\usepackage{amsmath}
\usepackage{amssymb}
\usepackage{graphicx}%
\begin{document}
\title{{\LARGE GRAVITOMAGNETISM AND ANGULAR MOMENTA OF BLACK-HOLES}}
\author{Marcelo Samuel Berman$^{1}$}
\affiliation{$^{1}$Instituto Albert Einstein/ Latinamerica\ - Av. Candido Hartmann, 575 -
\ \# 17}
\affiliation{80730-440 - Curitiba - PR - Brazil}
\begin{abstract}
We review the energy contents formulae of Kerr-Newman black-holes, where
gravitomagnetic energy term comes to play(Berman, 2006; 2006a; 2004). Then, we
obtain the angular momenta formulae, which include the gravitomagnetic effect.
Three theorems can be enunciated: (1) No black-hole has its energy confined to
its interior; (2) Rotating black-holes do not have confined angular momenta;
(3) The energy density of a black-hole is not confined to its interior.

The difference between our calculation and previous ones by Virbhadra(1990,
1990a, 1990b), and Aguirregabiria et al.(1996), lies in the fact that we
include a term responsible for the self-gravitational energy, while the cited
authors discarded such effect, which appears in the static black hole energy calculation.

(\textbf{Spanish}) Revisamos las formulas del contenido energetico de los
hoyos negros de Kerr-Newman, para los cuales la parte de energia
\ gravitomagnetica \ entra en escena.(Berman,2006;2006a;2004). Despues,
obtenemos las formulas de los momentos angulares, incluyendo el efecto
gravitomagnetico. Tres teoremas son establecidos: (1) Ningun hoyo negro tiene
su energia confinada en su interior; (2) Hoyos negros en rotacion no poseen su
momento angular confinado; (3) La densidad de energia de un hoyo negro no esta
confinada interiormente.

La diferencia entre nuestros resultados y los previos publicados por
Virbhadra(1990, 1990a, 1990b), y Aguirregabiria et al.(1996), se depreenden
por la ausencia de energia auto-gravitacional en el calculo efectuado por
aquellos autores. Esta ausencia se nota directamente en el calculo de la
energia de un hoyo negro est\'{a}tico.



\end{abstract}
\maketitle

\begin{center}
{\LARGE GRAVITOMAGNETISM AND ANGULAR MOMENTA OF BLACK-HOLES}

\bigskip

Marcelo Samuel Berman
\end{center}

\bigskip1. INTRODUCTION

\bigskip The calculation of energy and angular momentum of black-holes, has,
among others, an important \ astrophysical r\^{o}le, because such objects
remain the ultimate source of energy in the Universe, and \ the amount of
angular momentum is related to the possible amount of extraction of energy
from the b.h.(Levinson, 2006).

In a series of excellent papers, Virbhadra(1990; 1990a; 1990b) and
Aguirregabiria et al.(1996) calculated the energy contents, as well as the
angular momenta, for Kerr-Newman black-holes. Notwithstanding the high quality
of those papers, Berman(2004; 2006; 2006a) has pointed out that their results
for the energy do not reduce to the correct well known result by Adler et
al.(1975), when the electric charge and rotation parameters go to zero.
Furthermore, Berman(2004, 2006, 2006a) objected the energy formula (6) and (8)
below obtained by Virbhadra, and Aguirregabiria et al., because the
gravitomagnetic effect on the energy contents of the Kerr-Newman black-hole
does not appear in their results. Soon afterwards, Ciufolini and Pavlis(2004)
and Ciufolini(2005) reported the experimental verification of the
Lens-Thirring effect. This effect is a consequence of the concept of gravitomagnetism.

\bigskip

Therefore, it is now interesting to check whether the calculation of angular
momenta contents for a \ K.N. black hole given by \ Virbhadra, and
Aguirregabiria et al., includes the gravitomagnetic contribution. It will be
seen that this does not occur. We recalculate here the angular momenta
formulae, in order that gravitomagnetism enters into the scenario. We cite in
our favor, the papers by Lynden-Bell and Katz(1985) and Katz and Ori(1990).

2. CALCULATION OF ENERGY AND ANGULAR MOMENTA

\bigskip

The metric for a K.N. black hole may be given in Cartesian coordinates by:\ 

\bigskip$ds^{2}=dt^{2}-dx^{2}-dy^{2}-dz^{2}-\frac{2\left[  M-\frac{Q^{2}%
}{2r_{0}}\right]  r_{0}^{3}}{r_{0}^{4}+a^{2}z^{2}}\cdot F^{2}$ \ \ \ \ \ \ \ \ \ \ \ \ \ ,\ \ \ \ \ \ \ \ \ \ \ \ \ \ \ \ \ \ \ \ \ \ \ \ \ \ (1)

\bigskip while,

$F=dt+\frac{Z}{r_{0}}dz+\frac{r_{0}}{\left(  r_{0}^{2}+a^{2}\right)  }\left(
xdx+ydy\right)  +\frac{a\left(  xdy-ydx\right)  }{a^{2}+r_{0}^{2}}$\ \ \ \ \ \ \ \ \ \ \ ,\ \ \ \ \ \ \ \ \ \ \ \ \ \ \ \ \ \ \ \ \ \ \ \ \ \ (2)

\bigskip

$r_{0}^{4}-\left(  r^{2}-a^{2}\right)  r_{0}^{2}-a^{2}z^{2}=0$\ \ \ \ \ \ \ \ \ \ \ \ \ \ \ \ \ \ \ \ \ \ \ \ \ \ \ \ \ \ \ \ \ \ \ \ \ \ \ \ \ \ \ \ ,\ \ \ \ \ \ \ \ \ \ \ \ \ \ \ \ \ \ \ \ \ \ \ \ \ \ (3)

and,

\bigskip

$r^{2}\equiv x^{2}+y^{2}+z^{2}$ \ \ \ \ \ \ \ \ \ \ \ \ \ \ \ \ \ \ \ \ \ \ \ \ \ \ \ \ \ \ \ \ \ \ \ \ \ \ \ \ \ \ \ \ \ \ \ \ \ \ \ \ \ \ \ \ \ \ \ \ .\ \ \ \ \ \ \ \ \ \ \ \ \ \ \ \ \ \ \ \ \ \ \ \ \ (4)

\bigskip

In the above, \ $M$\ , \ $Q$\ \ \ and \ "$a$" \ stand respectively for the
mass, electric charge, and the rotational parameter, which has been shown to
be given by:

\bigskip

$a=\frac{J_{TOT}}{M}$\ \ \ \ \ \ \ \ \ \ \ \ \ \ \ \ \ , \ \ \ \ \ \ \ \ \ \ \ \ \ \ \ \ \ \ \ \ \ \ \ \ \ \ \ \ \ \ \ \ \ \ \ \ \ \ \ \ \ \ \ \ \ \ \ \ \ \ \ \ \ \ \ \ \ \ \ \ \ \ \ \ \ \ \ \ \ \ \ \ \ \ \ \ \ \ \ \ \ \ (5)

\bigskip

where \ $J_{TOT}$\ \ stands for the total angular momentum of the system.

\bigskip

According to Virbhadra and \ Aguirregabiria et al., the energy-momentum pseudo
tensor is given by:

\bigskip$\bar{P}_{0}=M-\left[  \frac{Q^{2}}{4\varrho}\right]  \left[
1+\frac{\left(  a^{2}+\varrho^{2}\right)  }{a\varrho}arctgh\left(  \frac
{a}{\varrho}\right)  \right]  $\ \ \ \ \ \ \ \ \ \ \ \ \ \ \ \ \ \ \ \ \ , \ \ \ \ \ \ \ \ \ \ \ \ \ \ \ \ \ \ \ \ \ \ \ \ \ \ \ \ \ \ (6)

\bigskip

\bigskip$P_{1}=P_{2}=P_{3}=0$\ \ \ \ \ \ \ \ . \ \ \ \ \ \ \ \ \ \ \ \ \ \ \ \ \ \ \ \ \ \ \ \ \ \ \ \ \ \ \ \ \ \ \ \ \ \ \ \ \ \ \ \ \ \ \ \ \ \ \ \ \ \ \ \ \ \ \ \ \ \ \ \ \ \ \ \ \ \ \ (7)

\bigskip

\bigskip It must be remarked that when we calculate, in this paper, energy or
angular momentum contents, we suppose that the calculation is done over a
closed surface with constant \ $\rho$\ \ ; the total energy and the total
angular momentum, should be calculated in the limit \ $\rho\rightarrow\infty$\ \ \ \ .\ 

\bigskip

If we expand relation (6) in powers of \ ($\frac{a}{\rho}$) , and retain only
the first and second powers, we obtain the approximate\ \ relation, valid for
slow rotational motion:\ 

\bigskip

$\bar{E}=\bar{P}_{0}\cong M-\left[  \frac{Q^{2}}{R}\right]  \left[
\frac{a^{2}}{3R^{2}}+\frac{1}{2}\right]  $\ \ \ \ \ \ \ \ \ \ \ \ \ \ \ \ .\ \ \ \ \ \ \ \ \ \ \ \ \ \ \ \ \ \ \ \ \ \ \ \ \ \ \ \ \ \ \ \ \ \ \ \ \ \ \ \ \ \ \ \ \ \ \ (8)

\bigskip

\noindent where $\varrho\rightarrow R$; this can be seen because the defining
equation for $\varrho$\ \ is:

$\frac{x^{2}+y^{2}}{\varrho^{2}+a^{2}}+\frac{z^{2}}{\varrho^{2}}=1$
\ \ \ \ \ and if \ \ \ $a\rightarrow$\ $0$, \ \ \ $\varrho\rightarrow
R$.\bigskip\ \ \ \ \ \ \ \ \ \ \ \ \ \ \ \ \ \ \ \ \ \ \ \ \ \ \ \ \ \ \ \ \ \ \ \ \ \ \ \ \ \ \ \ (9)

\bigskip

In the same token, the cited \ authors obtained, for angular momentum, defined by:

\bigskip

$J^{(3)}=\int\left[  x^{1}p_{2}-x^{2}p_{1}\right]  d^{3}x$ \ \ \ \ \ \ \ \ \ \ \ \ \ \ \ \ ,\ \ \ \ \ \ \ \ \ \ \ \ \ \ \ \ \ \ \ \ \ \ \ \ \ \ \ \ \ \ \ \ \ \ \ \ \ \ \ \ \ \ \ \ \ \ \ \ \ \ \ \ \ \ \ (10)

\bigskip

and for the above metric, \ \ 

\bigskip

$J^{(1)}=J^{(2)}=0$ \ \ \ \ \ \ \ , \ \ \ \ \ \ \ \ \ \ \ \ \ \ \ \ \ \ \ \ \ \ \ \ \ \ \ \ \ \ \ \ \ \ \ \ \ \ \ \ \ \ \ \ \ \ \ \ \ \ \ \ \ \ \ \ \ \ \ \ \ \ \ \ \ \ \ \ \ \ \ \ \ \ \ \ (11)

\bigskip

where \ $p_{i}$\ \ stand for the linear momentum densities ($i=1,2,3$)\ .

\bigskip

The cited authors also found:

\bigskip

$\bar{J}^{(3)}=aM-\left[  \frac{Q^{2}}{4\varrho}\right]  a\left[  1-\frac
{\rho^{2}}{a^{2}}+\frac{\left(  a^{2}+\varrho^{2}\right)  ^{2}}{a^{3}\varrho
}arctgh\left(  \frac{a}{\varrho}\right)  \right]  $\ \ \ \ \ \ \ \ \ , \ \ \ \ \ \ \ \ \ \ \ \ \ \ \ \ \ \ \ \ \ \ \ \ \ (12)

\bigskip

and when we go to the slow rotation case, Virbhadra found:

\bigskip

$\bar{J}^{(3)}\cong aM-2Q^{2}a\left[  \frac{a^{2}}{5R^{3}}+\frac{1}%
{3R}\right]  $ \ \ \ \ \ \ \ \ \ \ \ \ \ \ . \ \ \ \ \ \ \ \ \ \ \ \ \ \ \ \ \ \ \ \ \ \ \ \ \ \ \ \ \ \ \ \ \ \ \ \ \ \ \ \ \ \ \ \ \ \ \ \ \ (13)

\bigskip

Unfortunately, when \ \ $Q=0$\ \ in the above equations, we are left without
gravitomagnetic effects. The corrections made by Berman, in the above cited
papers, reside on the acknowledgment\ \ that due to the same \ $R^{-2}%
$\ \ dependence of the gravitation and electric interactions, as characterized
by Newton's law of gravitation, and Coulomb's law for electric charges, we
would have on a par, the equal contributions\ \ of charge and mass in the
above formulae, so that, except for the inclusion of the inertial term, we
should make the correction:

\bigskip

$Q^{2}\rightarrow Q^{2}+M^{2}$ \ \ \ \ \ \ \ \ \ \ . \ \ \ \ \ \ \ \ \ \ \ \ \ \ \ \ \ \ \ \ \ \ \ \ \ \ \ \ \ \ \ \ \ \ \ \ \ \ \ \ \ \ \ \ \ \ \ \ \ \ \ \ \ \ \ \ \ \ \ \ \ \ \ \ \ \ \ \ \ \ \ \ \ (14)

\bigskip

In corroboration of correction (14), we cite that, for the
Reissner-Nordstr\"{o}m metric, the energy formula is given by:

\bigskip

$E_{RN}=M-\left[  \frac{Q^{2}+M^{2}}{2R}\right]  $ \ \ \ \ \ \ \ \ \ \ \ . \ \ \ \ \ \ \ \ \ \ \ \ \ \ \ \ \ \ \ \ \ \ \ \ \ \ \ \ \ \ \ \ \ \ \ \ \ \ \ \ \ \ \ \ \ \ \ \ \ \ \ \ \ \ \ \ \ \ \ \ \ \ \ \ \ (15)

\bigskip

Relation(15) reduces correctly to the energy formula published by Adler et
al.(1975) for the spherical mass distribution, when we make \ $Q=0$\ \ in
(15). To the contrary, relations (6) and (8) do not reduce to Adler et al.'s
\ formula when \ $a=Q=0$\ \ , neither to the relation(15), when $a=0$\ \ in
relation(8).\ This shows that correction (14) \ is plausible. In fact, by
applying pseudo tensors, \ 

\bigskip

$P_{\mu}=%
{\displaystyle\int\limits_{t}}
\sqrt{-g}\left[  T_{\mu}^{0}+t_{\mu}^{0}\right]  d^{3}x=\ $constants\ \ \ \ \ \ \ ,\ \ \ \ \ \ \ \ \ \ \ \ \ \ \ \ \ \ \ \ \ \ \ \ \ \ \ \ \ \ \ \ \ \ \ \ \ \ \ \ \ \ \ \ \ \ \ (16)

where,

$\sqrt{-g}t_{\mu}^{V}=\frac{1}{2\complement}\left[  Ug_{\mu}^{\nu}%
-\frac{\partial U}{\partial g_{|\nu}^{\pi\beta}}g_{|\mu}^{\pi\beta}\right]  $\ \ \ \ \ \ \ \ \ \ \ \ \ \ \ \ \ ,\ \ \ \ \ \ \ \ \ \ \ \ \ \ \ \ \ \ \ \ \ \ \ \ \ \ \ \ \ \ \ \ \ \ \ \ \ \ \ \ \ \ \ \ \ \ (17)

$U=\sqrt{-g}g^{\rho\sigma}\left[  \left\{
\begin{array}
[c]{c}%
\begin{array}
[c]{c}%
\alpha\\
\sigma\varrho
\end{array}
\end{array}
\right\}  \left\{
\begin{array}
[c]{c}%
\begin{array}
[c]{c}%
\beta\\
\alpha\beta
\end{array}
\end{array}
\right\}  -\left\{
\begin{array}
[c]{c}%
\begin{array}
[c]{c}%
\alpha\\
\beta\varrho
\end{array}
\end{array}
\right\}  \left\{
\begin{array}
[c]{c}%
\begin{array}
[c]{c}%
\beta\\
\alpha\sigma
\end{array}
\end{array}
\right\}  \right]  $\ \ \ ,\ \ \ \ \ \ \ \ \ \ \ \ \ \ \ \ \ \ \ \ \ \ (18)

\bigskip

and,

$\complement=-\frac{8\pi G}{c^{2}}$\ \ \ \ \ \ \ \ \ \ \ \ \ \ \ \ , \ \ \ \ \ \ \ \ \ \ \ \ \ \ \ \ \ \ \ \ \ \ \ \ \ \ \ \ \ \ \ \ \ \ \ \ \ \ \ \ \ \ \ \ \ \ \ \ \ \ \ \ \ \ \ \ \ \ \ \ \ \ \ \ \ \ \ \ \ \ \ \ \ \ (19)

\bigskip

we find the correct relations for the energy and momentum:

\bigskip$P_{0}=M-\left[  \frac{Q^{2}+M^{2}}{4\varrho}\right]  \left[
1+\frac{\left(  a^{2}+\varrho^{2}\right)  }{a\varrho}arctgh\left(  \frac
{a}{\varrho}\right)  \right]  $\ \ \ \ \ \ \ \ \ \ \ \ \ \ \ ,\ \ \ \ \ \ \ \ \ \ \ \ \ \ \ \ \ \ \ \ \ \ \ \ (20)

$P_{1}=P_{2}=P_{3}=0$\ \ \ \ \ \ \ \ \ \ \ \ \ \ \ \ \ .\ \ \ \ \ \ \ \ \ \ \ \ \ \ \ \ \ \ \ \ \ \ \ \ \ \ \ \ \ \ \ \ \ \ \ \ \ \ \ \ \ \ \ \ \ \ \ \ \ \ \ \ \ \ \ \ \ \ \ \ \ \ \ \ \ \ (21)

\bigskip

\bigskip The last result "validates"\ the coordinate system chosen for the
present calculation: it is tantamount to the choice of a center-of-mass
coordinate system in Newtonian Physics, or the use of comoving observers in Cosmology.

By considering an expansion of the arctgh($\frac{a}{\varrho}$) function, in
terms of increasing powers of the parameter "$a$", and by neglecting terms
$a^{3}\simeq a^{4}\simeq....\simeq0$, we find the energy of a slowly rotating
Kerr-Newman black-hole,

$\bigskip$

$E=\bar{P}_{0}\simeq M-\left[  \frac{Q^{2}+M^{2}}{R}\right]  \left[
\frac{a^{2}}{3R^{2}}+\frac{1}{2}\right]  $\ \ \ \ \ \ \ \ \ \ \ \ \ \ \ \ \ \ \ \ \ ,\ \ \ \ \ \ \ \ \ \ \ \ \ \ \ \ \ \ \ \ \ \ \ \ \ (22)

\bigskip

\noindent where $\varrho\rightarrow R$ .

\bigskip

We can interpret the terms $\frac{Q^{2}a^{2}}{3R^{3}}$ \ and $\frac{M^{2}%
a^{2}}{3R^{3}}$ \ as the magnetic and gravitomagnetic energies caused by
rotation. Virbhadra(1990; 1990a; 1990b) and Aguirregabiria et al. (1996)
noticed the first of these effects in the year 1990, but since then it seems
that he failed to recognize the existence of the gravitomagnetic energy due to
$M$, on an equal footing.

\bigskip Likewise, if we apply:

\bigskip$J^{(3)}=\int\left[  x^{1}p_{2}-x^{2}p_{1}\right]  d^{3}x$ \ \ \ \ \ \ \ \ \ \ \ \ \ \ \ \ ,\ \ \ \ \ \ \ \ \ \ \ \ \ \ \ \ \ \ \ \ \ \ \ \ \ \ \ \ \ \ \ \ \ \ \ \ \ \ \ \ \ \ \ \ \ \ \ \ \ \ \ \ \ \ \ \ \ \ (10)

where, \ the linear momentum densities are given by:\ 

\bigskip

$p_{1}=-2\left[  \frac{\left(  Q^{2}+M^{2}\right)  \rho^{4}}{8\pi(\rho
^{4}+a^{2}z^{2})^{3}}\right]  ay\rho^{2}$ \ \ \ \ \ \ \ \ \ \ \ ,\ \ \ \ \ \ \ \ \ \ \ \ \ \ \ \ \ \ \ \ \ \ \ \ \ \ \ \ \ \ \ \ \ \ \ \ \ \ \ \ \ \ \ \ \ \ \ \ \ \ \ \ \ \ \ \ \ \ \ (23)

\bigskip

$p_{2}=-2\left[  \frac{\left(  Q^{2}+M^{2}\right)  \rho^{4}}{8\pi(\rho
^{4}+a^{2}z^{2})^{3}}\right]  ax\rho^{2}$ \ \ \ \ \ \ \ \ \ \ \ ,\ \ \ \ \ \ \ \ \ \ \ \ \ \ \ \ \ \ \ \ \ \ \ \ \ \ \ \ \ \ \ \ \ \ \ \ \ \ \ \ \ \ \ \ \ \ \ \ \ \ \ \ \ \ \ \ \ \ \ (24)

\bigskip

$p_{3}=0$ \ \ \ \ \ , \ \ \ \ \ \ \ \ \ \ \ \ \ \ \ \ \ \ \ \ \ \ \ \ \ \ \ \ \ \ \ \ \ \ \ \ \ \ \ \ \ \ \ \ \ \ \ \ \ \ \ \ \ \ \ \ \ \ \ \ \ \ \ \ \ \ \ \ \ \ \ \ \ \ \ \ \ \ \ \ \ \ \ \ \ \ \ \ \ \ \ \ \ \ \ \ (25)

\bigskip

while the energy density is given by:

\bigskip

$\mu=\left[  \frac{\left(  Q^{2}+M^{2}\right)  \rho^{4}}{8\pi(\rho^{4}%
+a^{2}z^{2})^{3}}\right]  \left(  \rho^{4}+2a^{2}\rho^{2}-a^{2}z^{2}\right)  $\ \ \ \ \ \ \ ,\ \ \ \ \ \ \ \ \ \ \ \ \ \ \ \ \ \ \ \ \ \ \ \ \ \ \ \ \ \ \ \ \ \ \ \ \ \ \ \ \ \ \ \ \ \ \ (26)

\bigskip

we find:

\bigskip

$J^{(3)}=aM-\left[  \frac{Q^{2}\text{ }+\text{ }M^{2}}{4\varrho}\right]
a\left[  1-\frac{\rho^{2}}{a^{2}}+\frac{\left(  a^{2}+\varrho^{2}\right)
^{2}}{a^{3}\varrho}arctgh\left(  \frac{a}{\varrho}\right)  \right]
$\ \ \ \ \ \ \ \ \ . \ \ \ \ \ \ \ \ \ \ \ \ \ \ \ (27)

\bigskip

\bigskip Expanding the \ $arctgh$ \ function in powers of ($\frac{a}{\varrho}%
$)\ , and retaining \ up to third power, we find the slow rotation angular momentum:

\bigskip

$J^{(3)}\cong aM-2\left[  Q^{2}+M^{2}\right]  a\left[  \frac{a^{2}}{5R^{3}%
}+\frac{1}{3R}\right]  $ \ \ \ \ \ \ \ \ \ \ \ \ \ \ . \ \ \ \ \ \ \ \ \ \ \ \ \ \ \ \ \ \ \ \ \ \ \ \ \ \ \ \ \ \ \ \ \ \ \ \ \ \ (28)\ \ 

\bigskip

In the same approximation, relation (26) would become:

\bigskip

$\mu\cong\left[  \frac{Q^{2}+M^{2}}{4\pi R^{4}}\right]  \left[  \frac{a^{2}%
}{R^{2}}+\frac{1}{2}\right]  $ \ \ \ \ \ \ \ \ \ \ \ . \ \ \ \ \ \ \ \ \ \ \ \ \ \ \ \ \ \ \ \ \ \ \ \ \ \ \ \ \ \ \ \ \ \ \ \ \ \ \ \ \ \ \ \ \ \ \ \ \ \ \ \ \ \ \ \ \ \ \ \ \ \ (29)

\bigskip

The above formula could be also found by applying directly the definition, \ \ \ \ 

\bigskip

$\mu=\frac{dP_{0}}{dV}=\frac{1}{4\pi R^{2}}\frac{dP_{0}}{dR}$ \ \ \ \ \ \ \ , \ \ \ \ \ \ \ \ \ \ \ \ \ \ \ \ \ \ \ \ \ \ \ \ \ \ \ \ \ \ \ \ \ \ \ \ \ \ \ \ \ \ \ \ \ \ \ \ \ \ \ \ \ \ \ \ \ \ \ \ \ \ \ \ \ \ \ \ \ \ \ \ \ (30)

\bigskip

where \ $P_{0}$\ \ would be given by the approximation (22), with \ $P_{0}%
=E$\ \ . (Berman, 2004; 2006; 2006a).

\bigskip3. FINAL COMMENTS AND CONCLUSIONS

The different approach in our paper, as compared with those of Virbhadra(1990,
1990a, 1990b), and Aguirregabiria et al.(1996), can be recognized from the
lack of a self-gravitational energy term, in those authors calculations; for
instance, Adler et al.(1975), calculated the self-gravitational-energy of a
spherical mass distribution, by the term \ $-\frac{GM^{2}}{2R}$\ \ .
\ However, we can not trace this term in their formulae (6) and (8); they are
present in our calculation, as in formulae (20) and (22). Except for the
inertial mass-energy term \ $M$\ \ , the self-gravitational and self-electric
energies, in our calculation, present similar contributions, which, for the
static black hole, are given by \ ( $-\frac{GM^{2}}{2R}$\ )\ \ and \ \ (
$-\frac{Q^{2}}{2R}$\ \ )\ . This means that where those authors worked only
with an electric term \ \ $-\frac{Q^{2}}{2R}$\ \ \ , \ we must work with both
mass and charge contributions.

\bigskip

We recollect now a series of statements that we have shown above to be
incorrect, and which appeared in the papers by Virbhadra and Aguirregabiria et al.:

A) no angular momentum is associated with the exterior in Kerr's metric;

B) no energy is shared by the exterior of the Kerr black hole;

C) the energy density in the Kerr black hole equals zero;

D) the energy density in the Schwarzschild's black hole equals zero;

E) the entire energy of Schwarzschild's black hole is confined to its interior.

\bigskip

Instead, three correct statements are issued by us:

(1) No black-hole has its energy confined to its interior;

(2) Rotating black-holes do not have confined angular momenta;

(3) The energy density of a black-hole is not confined to its interior.

\bigskip

We further conclude that we may identify the gravitomagnetic contribution to
the energy and angular momentum of the K.N. black hole, for the slow rotating
case, as:

\bigskip

$\Delta E\cong-\frac{M^{2}a^{2}}{3R^{3}}$ \ \ \ \ \ \ , \ \ \ \ \ \ \ \ \ \ \ \ \ \ \ \ \ \ \ \ \ \ \ \ \ \ \ \ \ \ \ \ \ \ \ \ \ \ \ \ \ \ \ \ \ \ \ \ \ \ \ \ \ \ \ \ \ \ \ \ \ \ (31)

\bigskip

and,

$\Delta J\cong-2M^{2}\left[  \frac{a^{3}}{5R^{3}}+\frac{a}{3R}\right]
\approx-\frac{2M^{2}a}{3R}$ \ \ \ \ , \ \ \ \ \ \ \ \ \ \ \ \ \ \ \ \ \ \ \ \ \ \ \ \ \ \ \ \ \ \ \ \ (32)

\bigskip

as can be checked from relations (8) and (28) above.

\bigskip

\ \ \ 

{\Large Acknowledgements}

\bigskip

\bigskip The author was saved from unclear passages by an anonymous referee to
whom I am grateful, as well as with my intellectual mentors, Fernando de Mello
Gomide and M. M. Som, and I am also grateful for the encouragement by Paula,
Albert, and Geni and also to Marcelo Guimar\~{a}es, Nelson Suga, Mauro
Tonasse, Antonio Teixeira and Herman J. M. Cuesta.

\bigskip

{\Large References}

\bigskip

Adler, R.; Bazin, M,; Schiffer, M. (1975) - \textit{Introduction to General
Relativity}\ - 2nd. edtn., McGraw-Hill, N.Y.

Aguirregabiria, J.M. et al. (1996) - Gen. Rel. and Grav. \textbf{28}, 1393.

Berman,M.S. (2004) - gr-qc/0407026

Berman,M.S. (2006) - \textit{Energy of Black-Holes and Hawking's Universe
\ }in \textit{Trends in Black-Hole Research, }Chapter 5\textit{.} Edited by
Paul Kreitler, Nova Science, New York.

Berman,M.S. (2006 a) - \textit{Energy, Brief History of Black-Holes, and
Hawking's Universe }in \textit{New Developments in Black-Hole Research},
Chapter 5\textit{.} Edited by Paul Kreitler, Nova Science, New York.

Ciufolini, I. (2005) - gr-qc/0412001 v3

\noindent Ciufolini, I.; Pavlis, E. (2004) - Letters to Nature, \textbf{431, }958.

Ciufolini, I.; Wheeler, J. A. (1995) - \textit{Gravitation and Inertia},
Princeton Univ. Press, Princeton. See especially page 82, \# 72 to 83.

Katz, J.; Ori, A. (1990) - Classical and Quantum Gravity \textbf{7}, 787.

Levinson, A. (2006) - in \textit{Trends in Black Hole Research,} ed. by Paul
Kreitler, Nova Science, New York.

Lynden-Bell,D.;Katz, J. (1985) - M.N.R.A.S.\textbf{ 213},21.

Virbhadra, K.S. (1990) - Phys. Rev. \underline{\textbf{D41}}, 1086.

Virbhadra, K.S. (1990a) - Phys. Rev. \underline{\textbf{D42}}, 2919.

Virbhadra, K.S. (1990b) - Phys. Rev. \underline{\textbf{D42}}, 1066.

\end{document}